# The human longevity record may hold for decades

Jeanne Calment's extraordinary record is <u>not</u> evidence for an upper limit to human lifespan

Adam Lenart[1], José Manuel Aburto[1,3], Anders Stockmarr[2], James W. Vaupel[1,3*]

**1** Max-Planck Odense Center on the Biodemography of Aging, Department of Epidemiology, Biostatistics and Biodemography, University of Southern Denmark, Odense, Denmark
**2** Department of Applied Mathematics and Computer Science, Technical University of Denmark, Kongens Lyngby, Denmark
**3** Max Planck Institute for Demographic Research, Rostock, Germany
* address correspondence to jwv@demogr.mpg.de

**Abstract**

Since 1990 Jeanne Louise Calment has held the record for human longevity. She was born on 21 February 1875, became the longest-lived human on 12 May 1990 when she was 115.21 and died on 4 August 1997 at age 122.45 years. In this chapter, we use data available on 25 September 2017 on people who reached age 110, supercentenarians, to address the following questions:
(1) How likely is it that a person has reached age 122.45?
(2) How unlikely is it that Calment's record has not yet been broken?
(3) How soon might it be broken?
Assuming a constant annual probability of death of 50% after age 110, we found that the probability that a person who survived to age 110 would have lived to 122.45 by 25 September 2017 is 17.1%. Furthermore, we calculated that there was only a 20.3% chance that Calment's record would have been broken after 1997 but before 2017. Finally, we estimated that there is less than a 50% chance that someone will surpass Calment's lifespan before 2045. Jeanne Louise Calment's record is exceptional but not impossible. It does not provide evidence that her lifespan is an upper limit to human lifespans.

Jeanne Louise Calment lived 122.45 years, longer than any other reliably documented person. She was born in Arles, France on 21 February 1875 and died there on 4 August 1997. Her date of death and that the person who died was indeed the person born 122.45 years earlier have been meticulously verified *(1,2)*. Her record seems remarkable because it is so high and because it has not yet been broken. This has been taken as evidence that human longevity is approaching a biological limit *(3,4)*. Such a ceiling would have important implications for research policy since a ceiling implies that extension of human longevity requires unprecedented research breakthroughs that overcome the physiology of aging. Hence it is of interest to ask: (1) How likely is it that a person reached age 122.45? (2) How unlikely is it that this record has not yet been broken? (3) How soon might it be broken?

Here we exploit data on the number of supercentenarians (people who reach age 110). We used the data to estimate the probability distribution of the age of the record holder on 25 September 2017. To assess the probability that Calment's record would be broken by this date, we counted the number of supercentenarians alive when she died as well as the number of subsequent supercentenarians who could have reached the age of 122.45 by 25 September 2017. We forecast the time when a supercentenarian would reach an age greater than 122.45 by using observed data on supercentenarians and predicting their future numbers.

Three lists of supercentenarians are available. One is compiled by a research team working on the International Database on Longevity (IDL) *(5 and this book)*. This list only includes supercentenarians in countries for which members of the research team have undertaken careful verification and for which exhaustive lists of supercentenarians are available. Two other lists, one compiled by Louis Epstein and colleagues and the other by Robert Douglas Young and colleagues at the Gerontology Research Group (GRG), include verified supercentenarians from any country *(6)*.

We used the GRG list to estimate how many people reached age 110 over time, because this list is more inclusive than the IDL list and appears to be more reliable than the Epstein list. We relied on the IDL list, however, for information on the age-pattern of mortality after age 110. The IDL list was created such that the probability of the inclusion of a person does not depend on the person's age; the list includes all age-validated supercentenarians in the populations studied. The inclusion criteria for the GRG and Epstein lists are not as strict as those of the IDL: some individuals might be accepted to the lists because of attracting public attention, perhaps via a newspaper article. The probability of attention tends to increase with age; for example, 115-year-olds are more likely to be included on the lists than a person who dies shortly after her or his 110[th] birthday.

Previous research on supercentenarians based on the IDL database *(7)* and subsequent research published in this book show that between ages 110 and 114 the annual probability of death is close to 50%. Data are too sparse after 114 to estimate risks of mortality reliably, but it may be reasonable to assume that the probability of dying reaches a plateau at a level of 0.5/year *(8,9)*. GRG data are consistent with the probability of death levelling off at about 50%. Furthermore, if the probability of death per year is assumed to be constant after age 110, then the GRG data indicate that this probability is 0.491 with 95% confidence interval 0.474-0.508.

It is known that Jeanne Calment reached age 122.45 in 1997. Hence the likelihood of this is 100%. To estimate a probability for this record we have to put ourselves behind a veil of ignorance. One way to do so is as follows.

According to the Young list, 1,049 people celebrated their 110[th] birthday more than 12.45 years before 25 September 2017. Assuming a 50% annual probability of death after age 110, the probability that one of them would have reached age 122.45 is 17.1% (see BOX). Jeanne Calment's age at death is extraordinary but not impossibly so.

Now consider how extreme it is that Calment's record has not yet been broken. We know it has not been broken, so as above we have to put ourselves behind a veil of ignorance. An appealing way to do so is to estimate the probability, on the day of her death, that at least one supercentenarian would exceed her record before 25 September 2017. Her record could have been broken either by supercentenarians who were alive when she died on 4 August 1997 or by new supercentenarians who attained age 110 following her death but at least 12.45 years before 25 September 2017, permitting them to reach age 122.45 by this date.

Young's list shows that 67 supercentenarians were alive on the day of Calment's death. We now know the number of new supercentenarians afterwards but this was not known when Jeanne Calment died. The new group of supercentenarians needs to be estimated from the data potentially available on her day of death. Over the 10-year period preceding her demise, annual counts of newly recruited supercentenarians grew on average by 8.0% annually. Extending this trend line until 12.45 years before 25 September 2017 yields 742 new supercentenarians. If we combine analysis of the 67 supercentenarians whose exact ages were known with analysis of the 742 people who reached age 110 afterwards, we arrive at a probability of 20.4% that at least one of them would reach the age of 122.45 (see BOX). This is an estimate of the chance that Calment's record would have been broken after 1997 but before 2017. It is not surprising that her record still holds.

It is reasonable to predict when Calment's record might be broken based on the available data, i.e., to make the prediction considering the fact that her record has not been broken by 25 September 2017.

The number of newly registered supercentenarians on the Young list dropped sharply after 2010, probably because of delays in registering and validating alleged new cases. On the list available on 25 September 2017, there were no 110 or 111-year-olds; only one person on the list at the end of 2010 was still alive, at age 117. Because the list available on 25 September 2017 is problematic, we decided not to use any of the information in it, including the information that there was one 117-year-old. Instead, we decided to base our estimates on data available for the period up until the end of 2010. It turns out that applying the 0.5 annual probability of survival to this 2010 list leads to the prediction that there would be one 117-year-old on 25 September 2017, which made us more comfortable about neglecting this information from the 25 September 2017 list. We used the GRG data to determine the number of people who celebrated their 110[th] birthdays over the period beginning on 25 September 2017 minus 12.45 years and ending on 31 December 2010; the number is 165. Then, using 77, the number of people who attained age 110 in 2010, as our base value, we estimated the number of new supercentenarians in subsequent years. We did so by using a continuous growth rate of 6.8%/year, which leads to annual growth by multiples of 1.07 (because exp(0.068)=1.07). Hence we calculated the cumulative sum of 1.07 times 77 plus 1.07 squared times 77 plus .... Calment's record is likely to be broken when the probability of someone living longer than 122.45 exceeds 50%. To reach this level, we need to mix the estimated number of people attaining age 110 in the period prior to the end of 2010 (165 people) with 3,713 new supercentenarians. The required number of new supercentenarians would not be reached until near the end of 2033. Then it would take an additional 12.45 years for one of them to break Calment's record, so about mid-year 2045, assuming the annual probability of death after age 110 remains at 0.5.

Is it plausible that Jeanne Calment might hold the longevity record from 23 May 1989, when she reached age 115.21 and broke the previous record, until 2045, some 56 years later? The Table shows how long previous records endured, for males as well as for females. Delina Filkins became the world's longest-lived woman in 1926 when she attained age 111.38; her record held until the end of 1980, over 54 years later. Among men, Gert Boomgaard died at age 110.37 in 1899; after his death his record lasted more than 67 years until 1966. Extreme-value theory is far from intuitive; jumps in maximum values are often surprising *(10)*.

Table: List of Oldest Females and Males (up to 25 September 2017)

|  |  | Name | Age at Death | Became Oldest | Date of death | Years held record |
|---|---|---|---|---|---|---|
| Females | 1 | Margaret Ann Neve | 110.85 | ? | 1903-04-04 | >21 |
|  | 3 | Delina Filkins | 113.59 | 1926-03-21 | 1928-12-04 | 54.66 |
|  | 4 | Fannie Thomas | 113.75 | 1980-11-15 | 1981-01-22 | 4.49 |
|  | 5 | Augusta Holtz | 115.21 | 1985-05-14 | 1986-10-21 | 4.99 |
|  | 6 | Jeanne Calment | 122.45 | 1990-05-12 | 1997-08-04 | 27.37 |
| Males | 1 | Gert Adrians-Boomgaard | 110.37 | ? | 1899-02-03 | >67 |
|  | 2 | John Turner | 111.77 | 1966-10-29 | 1968-03-21 | 14.46 |
|  | 3 | Mathew Beard | 114.61 | 1981-04-16 | 1985-02-16 | 16.03 |
|  | 4 | Christian Mortensen | 115.69 | 1997-04-26 | 1998-04-25 | 15.67 |
|  | 5 | Jiroemon Kimura | 116.15 | 2012-12-28 | 2013-06-12 | 4.74 |

Source: (6)

Jeanne Calment's exceptional lifespan provides evidence about another question. The force of mortality (hazard of death) increases approximately exponentially after age 50: the pattern is known as Gompertz' law. Strong evidence from various studies shows that the force of mortality levels off at advanced ages *(5,8)*. It has been claimed, however, that Gompertz law continues to hold up to the most advanced ages *(11)*. If the rate of increase in mortality after 110 is similar to the rate of increase at younger ages, then the age reached by Jeanne Calment is incompatible with this claim. Suppose the force of mortality continued to increase after age 110 at the rate of 10% per year observed at younger ages and suppose the increase at younger ages resulted in a probability of death at age 110 of 50%. Then, the probability an individual could survive to 122.45 would be 0.00000008, three orders of magnitude less than then the 0.00018 chance if the force of mortality after 110 reached a plateau. The probability of a lifespan of 122.45 in our era would be minuscule. More generally, the survival of hundreds of individuals beyond age 110 is consistent with a marked slowing of death rates after age 100 and probably a plateau of mortality after age 110 or even earlier (as suggested in other chapters in this book).

Jeanne Louise Calment is a truly exceptional person but not an impossible one. It is within the realm of possibility that someone could survive to age 122.45. It is not surprising that her record stands. It is likely that her record will not be broken for decades. She should not be used to argue that human lifespans are approaching ultimate limits. Progress to date in increasing lifespans has been due to large reductions in death rates *(12)*; progress since 1950 has been due to the postponement of high levels of mortality to higher ages *(8)*. It is possible (albeit uncertain) that this long-term trend will continue: Jeanne Calment's enduring longevity record does not provide evidence to the contrary.

**BOX**

Let $p$ denote the probability that a person attaining age 110 reaches age 122.45. Then $1 - p$ is the chance the person will die before age 122.45. Assuming that the risk of death is 50%/year after age 110, the chance of reaching 110 is $0.5^{12.45}$, which is approximately 0.00018 or about 1 in 5595. The corresponding chance that a person who reaches 110 will not reach 122.45 is $1-0.5^{12.45}$ or about 99.982%.

If $N$ people attain age 110, then the probability that all of them will die before age 122.45 is

$$(1-p)^N.$$

To calculate the probability that at least one person reaches age 122.45, this probability has to be subtracted from 1:

$$1 - (1-p)^N.$$

Hence, the probability that at least one person out of N people will survive from 110 to at least 122.45 is

$$1 - [1 - 0.5^{12.45}]^N,$$

which equals 17.1% if $N$ = 1,049.

If a person is age $x > 110$, the probability that this person would reach 122.45 is given by $0.5^{122.45-x}$. The chance that none of a group of $n$ individuals with ages $x_1, x_2, \ldots, x_n$ would reach age 122.45 is given by:

$$\prod_{i=1}^{n}(1 - 0.5^{122.45-x_i}).$$

If a population is being analyzed that includes $n$ people whose age is greater than 110 and $N$ people whose age is exactly 110, then the probability that at least one of them will reach age 122.5 is:

$$1 - \left\{\prod_{i=1}^{n}(1 - 0.5^{122.5-x_i})\right\}(1 - 0.5^{12.45})^N.$$

**ACKNOWLEDGMENTS**

The authors thank Trifon Missov and Jim Oeppen for helpful comments, Robert D. Young for sharing the data that he and other members of the Gerontology Research Group collected, and Nicolas Brouard for pointing out that Calment's record is inconsistent with a continuing exponential increase in mortality at advanced ages.